\documentclass[12pt]{iopart}
\usepackage{epsfig}
\usepackage{iopams}  
\begin{document}

\title{No Black Holes at IceCube}

\author{Ulrich Harbach}

\address{Institut f{\"u}r Theoretische Physik\\
Johann Wolfgang Goethe-Universit{\"a}t\\
and\\
Frankfurt Institute for Advanced Studies\\
Max-von-Laue-Str. 1\\
60438 Frankfurt am Main, Germany}
\ead{harbach@th.physik.uni-frankfurt.de}

\author{Marcus Bleicher}
\address{Institut f{\"u}r Theoretische Physik\\
Johann Wolfgang Goethe-Universit{\"a}t\\
Max-von-Laue-Str. 1\\
60438 Frankfurt am Main, Germany} 
\ead{bleicher@th.physik.uni-frankfurt.de}

\begin{abstract}
We discuss modifications of the neutrino-nucleon cross section due to a minimal length scale. 
It is shown that the enhancement of the $\nu$-N cross section due to new physics is suppressed. 
Especially the potential observation rate of micro black holes at neutrino telescopes is strongly reduced.
\end{abstract}

\maketitle

Ultra High Energy Cosmic Rays (UHECR) contain the highest energetic particles we can observe today. 
Thus, they provide a unique possibility to test theories beyond the standard model, see e.g. \cite{Domokos:1998ry,Emparan:2001kf,Ringwald:2001vk,Anchordoqui:2001cg,Kazanas:2001ep,Kowalski:2002gb,Alvarez-Muniz:2002ga,Sigl:2002bb,Anchordoqui:2003jr,Anchordoqui:2004ma,Anchordoqui:2005gj,Anchordoqui:2005is}. Since cosmic rays consist mostly of neutrinos, the measurement of interactions of ultra high energy neutrinos with nucleons has received much attention, and many experiments such as Amanda, Auger and soon IceCube detect signatures of these interactions.

The string theory-motivated model of large extra dimensions\cite{Arkani-Hamed:1998rs,Antoniadis:1998ig} predicts -among else- a strong increase of the ultra high energy neutrino-nucleon cross section due to new physics, e.g. black hole production, which would result in horizontal air showers\cite{Emparan:2001kf,Anchordoqui:2001cg,Ringwald:2001vk,Feng:2001ib,Jain:2000pu,Anchordoqui:2001ei,Ahn:2003qn,Ave:2003ew}. In fact, the nonobservation of these showers has lead to constraints on the cross section and thus, on the parameters of the model\cite{Tyler:2000gt,Anchordoqui:2003jr,Anchordoqui:2005ey}. In addition, black hole events could be detected directly in a subsurface detector\cite{Uehara:2001yk}, and substantial rates have been predicted in realistic calculations\cite{Kowalski:2002gb}.

Recently it has been argued that stringent bounds on the parameters of the large extra dimensions model are questionable due to uncertainties in the cosmic neutrino flux and the black hole production cross section\cite{Ahn:2003cz}. In addition, effects due to a minimal length scale, which is also predicted by string theory, strongly question the validity of the semiclassical approach to the black hole production cross section at low energies and thus increase the minimal mass of black holes produced\cite{Cavaglia:2003qk,Cavaglia:2004jw}. Furthermore, these effects also substantially decrease high energy cross sections in general, and thus also the black hole production cross section\cite{Hossenfelder:2004ze}. In this paper, we examine the influence of these minimal length effects on the black hole production rates in subsurface neutrino detectors.

The remainder of the paper is organised as follows: First, we will review the standard model $\nu$-N cross section and the possibility of black hole production that arises in models with large extra dimensions. Then, we will introduce a toy model to include effects due to a minimal length scale and exploit the impact of such a notion on the various contributions to the total $\nu$-N cross section. Finally, we elaborate our results into realistic predictions for neutrino detectors, as an example, we calculate the black hole production rates for IceCube.

\section{The $\nu$-N cross section in Large Extra Dimensions}

In the Standard Model, the neutrinos only interact weakly with other particles, that is, via exchange of a W or Z boson. Within the electroweak model, the charged current differential cross section for scattering of a neutrino with an isoscalar nucleon $N=(p+n)/2$ can be written in terms of the Bjorken scaling variables $x=Q^2/2M\nu$ and $y=\nu/E_\nu$ as
\begin{equation}\label{sigmanun}
\frac{d^2\sigma}{dxdy}=\frac{2G_F^2M_NE_\nu}{\pi}\bigg(\frac{M_W^2}{Q^2+M_W^2}\bigg)^2\big(xq(x,Q^2)+x\bar{q}(x,Q^2)(1-y)^2\big)\quad.
\end{equation}
Here, $G_F=1.16632 \times 10^{-5} {\rm GeV}^{-2}$ is the Fermi constant, $M_W$ and $M_N$ are the weak boson and nucleon masses, $E_\nu$ is the neutrino energy in the nucleon rest frame, $Q^2$ and $\nu$ are the transferred momentum and energy, and $q(x,Q^2)$, $\bar{q}(x,Q^2)$ are the quark and antiquark distribution functions for an isoscalar nucleon. For details, see e.g. \cite{Gandhi:1995tf}. For numerical calculations, we use the CTEQ6 parton distribution functions.

In models with Large Extra Dimensions\cite{Arkani-Hamed:1998rs,Antoniadis:1998ig}, the observed weakness of gravity compared to the other fundamental forces (and thus, the hugeness of the Planck mass $M_{Pl}$) is only a consequence of the size of $d$ extra spatial dimensions. The fundamental mass scale $M_f^{2+d}=M_{Pl}^2/R^d$ of gravity can be as low as the electroweak symmetry breaking scale.\footnote{Here, $R^d$ is the volume of the compactified space. Note that there are various definitions of a new fundamental scale in literature, depending on the way of compactification.} Accordingly, at distances below the size of the extra dimensions, gravitation gains strength and black hole formation becomes possible on small scales\cite{Argyres:1998qn} due to the enormous rise of the Schwarzschild radius in higher dimensions\cite{Myers:1986un}
$$r_S(M_{BH})=\bigg (\frac{8\Gamma(\frac{d+3}{2})}{\sqrt{\pi}(d+3)}\frac{M_{BH}}{M_f^{d+2}}\bigg )^\frac{1}{d+1}\quad .$$
The cross section for black hole production for two point particles can be estimated on geometrical grounds and is given by
$$\sigma_{ij\rightarrow BH}\big(\sqrt{s}\big)=\pi r_S^2(\sqrt{s}\big)\theta(\sqrt{s}-M_{min}\big)\quad.$$
where $\sqrt{s}$ is the total energy of the colliding particles in the center-of-mass frame and $\theta$ is the Heaviside step function which provides a threshold $M_{min}$ for black hole production, i.e. an energy scale above which one sufficiently believes in the semiclassical picture adopted here.

For $\nu$-N interactions, where the nucleon has of course to be treated as a compound object, the cross section reads
\begin{equation}\label{sigmaBH}
\sigma_{\nu N\rightarrow BH}(\sqrt{s})=\int_0^1{\rm d}x \pi r_S^2(\sqrt{xs})\theta(\sqrt{xs}-M_{min})[q(x,\mu)+\bar{q}(x,\mu)]\quad,
\end{equation}
where $s=2M_NE_\nu$. For the factorisation scale of the parton distribution functions, we use the canonical choice $\mu=1/r_S$.

\section{The minimal length}

The necessity of more than 3 spatial dimensions is not the only prediction of string theory. In perturbative string theory\cite{Gross:1988ar,Amati:1988tn}, the feature of a fundamental minimal length scale arises from the fact that strings cannot probe distances smaller than the inverse string scale. If the energy of a string reaches this scale $M_s=\sqrt{\alpha'}$, excitations of the string can occur and increase its extension\cite{Witten:1997fz}. In particular, an examination of the spacetime picture of high-energy string scattering shows that the extension of the string is proportional to its energy\cite{Gross:1988ar} in every order of perturbation theory. Due to this, uncertainty in position measurement can never become arbitrarily small.

To include effects of the minimal length, we use the model developed in \cite{Hossenfelder:2003jz,Hossenfelder:2004up}. It is assumed that at arbitrarily high momentum $p$ of a particle, its wavelength is bounded by some minimal length $L_{\mathrm f}$ or, equivalently, its wave-vector $k$ is bounded by a $M_{\mathrm f}=1/L_{\rm f}$\cite{Ahluwalia:2000iw}. Thus, the relation between the momentum $p$ and the wave vector $k$ is no longer linear $p=k$ but a function $k=k(p)$\footnote{Note, that this is similar to introducing an energy dependence of Planck's constant $\hbar$.}, which is strongly constrained by the following properties:
\begin{enumerate}
\item[a)]  For energies much smaller than the new scale it yields the linear relation:
for $p \ll M_{\mathrm f}$ we have $p \approx k$. \label{limitsmallp}
\item[b)] It is an an uneven function (because of parity) and $k \parallel p$.
\item[c)]  The function asymptotically approaches the bound $M_{\mathrm f}$. \label{upperbound}
\end{enumerate}
The quantisation in this scenario is straightforward and follows the usual procedure. Using the well known commutation relations
$$[\hat x_i,\hat k_j]={\mathrm i } \delta_{ij}\quad$$
and inserting the functional relation between the wave vector and the momentum then yields the modified commutator for the momentum and results in the generalized uncertainty principle ({\sc GUP})\cite{Kempf:1995su}
\begin{equation} \label{CommXP}
[\,\hat{x}_i,\hat{p}_j]= + {\rm i} \frac{\partial p_i}{\partial k_j} \quad\longrightarrow\quad \Delta p_i \Delta x_j \geq \frac{1}{2}  \Bigg| \left\langle \frac{\partial p_i}{\partial k_j}\right\rangle \Bigg| \quad,
\end{equation}
which reflects the fact that it is not possible to resolve space-time distances arbitrarily well. Because $k(p)$ becomes asymptotically constant, its derivative $\partial k/ \partial p$ eventually vanishes and the uncertainty (Eq.(\ref{CommXP})) increases for high momenta. Thus, the introduction of the minimal length through this model reproduces the limiting high energy behavior found in string theory\cite{Gross:1988ar}.

The arising physical modifications can be traced back to an effective replacement of the usual momentum measure by a measure which is suppressed at high momenta:
$$\frac{{\mathrm d}^{3} p}{(2 \pi)^{3}} \rightarrow \frac{{\mathrm d}^{3} p}{(2 \pi)^{3}}
\Bigg| \frac{\partial k}{\partial p}
\Bigg|  \quad, \label{rl}$$
This replacement is founded by the finiteness of the integration bounds in $k$-space. Here, the absolute value of the partial derivative denotes the Jacobian determinant of $k(p)$. 

For the calculations in section \ref{nunml}, we will use the specific relation from \cite{Hossenfelder:2004up} for $k(p)$, i.e. the error function
$$k_{\mu}(p) = \hat{e}_{\mu} \int_0^{p} e^{{\displaystyle{-\epsilon p'^2}}} {\rm d} p'\label{model} \quad,$$
where $\hat{e}_{\mu}$ is the unit vector in $\mu$-direction, $p^2=\vec{p}\cdot\vec{p}$ and
$\epsilon=L_{\mathrm f}^2 \pi / 4 $ (the factor $\pi/4$ is included to assure, that the limiting value is indeed $1/L_{\mathrm f}$).
It is easily verified that this expression fulfills the requirements (a) - (c).

\section{Neutrino interactions with minimal length}\label{nunml}

\begin{figure}
\includegraphics[width=13cm]{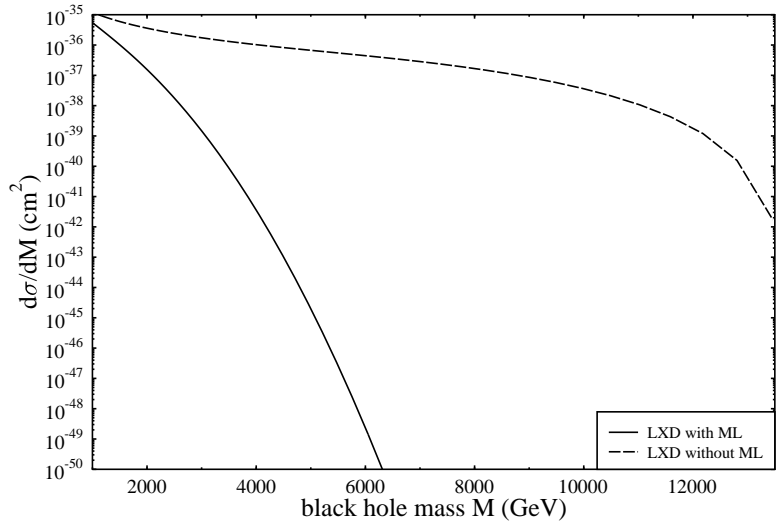}
\caption{Mass distribution of black holes produced in $\nu$-N interaction with $E_\nu=10^8$~GeV. The dashed line indicates the standard calculation, the solid line shows the result with minimal length included. Here, $M_f=1 {\rm TeV}$, $d=6$, $M_{min}=M_f$.\label{fig1}}
\end{figure}

\begin{figure}
\includegraphics[width=13cm]{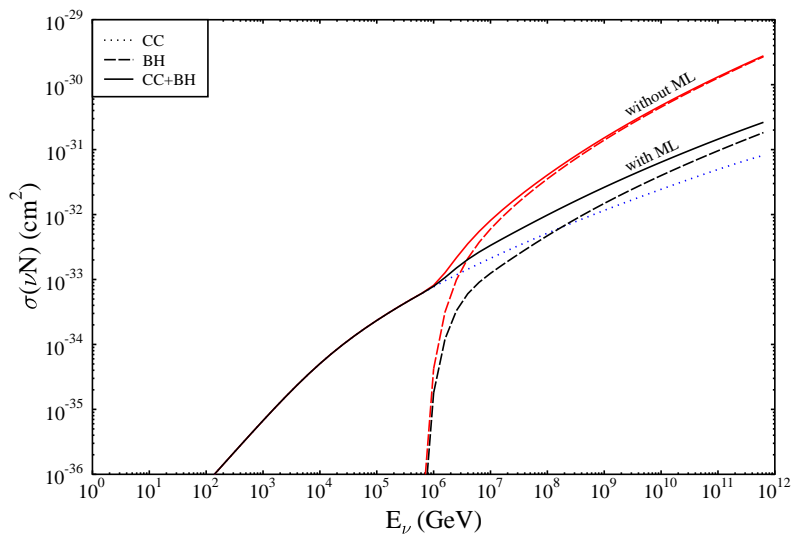}
\caption{Total $\nu$-N cross section as a function of the incident neutrino energy. The dotted line depicts the charged current cross section, the dashed lines depict the contribution from black hole production and the solid lines yield the respective sums. Here, $M_f=1 {\rm TeV}$, $d=6$, $M_{min}=M_f$.\label{fig2}}
\end{figure}

\begin{figure}
\includegraphics[width=14cm]{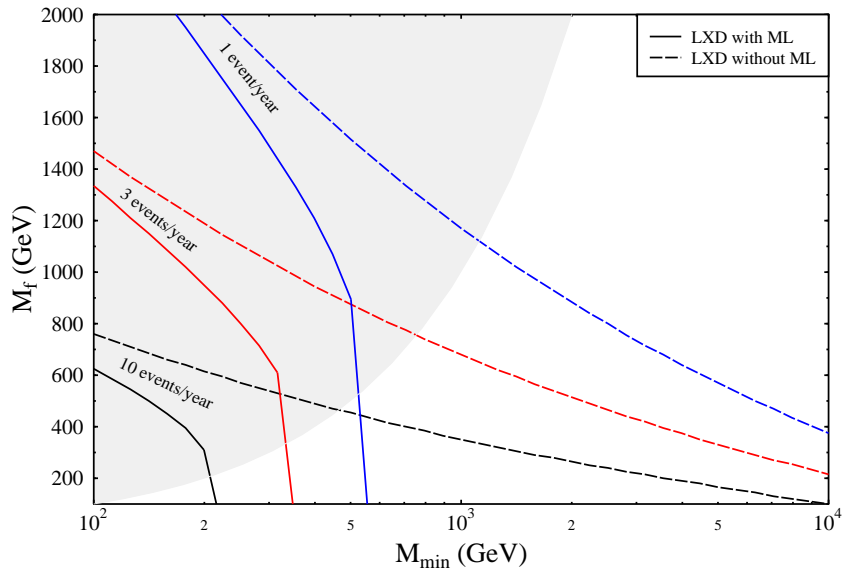}
\caption{Black hole production rates contained in a volume of 1÷km$^3$ at 2÷km below the surface as a function of the fundamental scale and the minimal mass of the black holes. Here, a conservative estimate of the cosmogenic neutrino flux from Ref. \cite{Protheroe:1995ft} is assumed. The shaded region is excluded ($M_{min} < M_f$). Again, $d=6$.\label{fig3}}
\end{figure}

\begin{figure}
\includegraphics[width=14cm]{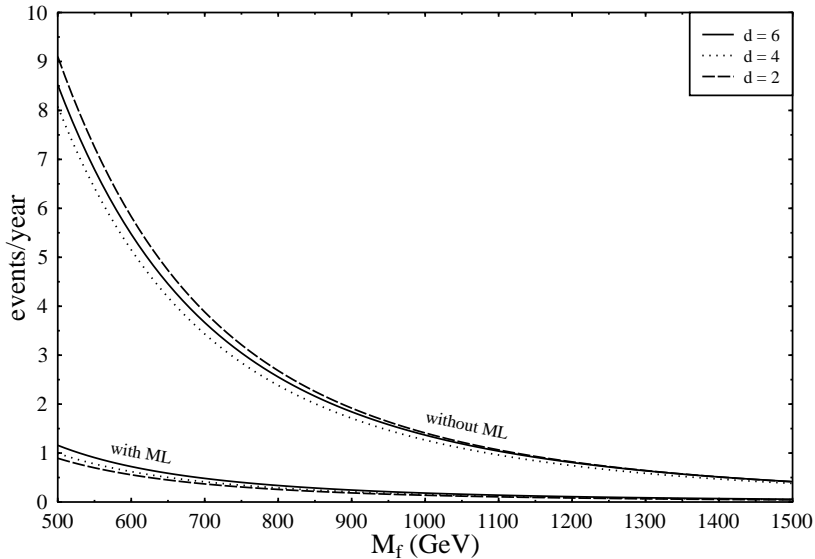}
\caption{Black hole production rates contained in a volume of 1~km$^3$ at 2~km below the surface as a function of the fundamental scale for different numbers of extra dimensions. The upper curves are calculated without minimal length, while the lower curves show the calculations with the minimal length included. Here, a conservative estimate of the cosmogenic neutrino flux from Ref. \cite{Protheroe:1995ft} is assumed, $M_f=M_{min}=1$~TeV.\label{fig4}}
\end{figure}

The new momentum space measure has a direct impact on the $\nu$-N cross sections (\ref{sigmanun}) and (\ref{sigmaBH}). Interactions with partons that carry a high momentum fraction are suppressed in models with a minimal length \cite{Hossenfelder:2004ze}. The results are depicted in Figs. \ref{fig1} and \ref{fig2}: Fig. \ref{fig1} shows that the mass distribution of black holes is strongly diminished in the high mass region, which leads to a global decrease of the total black hole production cross section. 

Fig. \ref{fig2} shows the total $\nu-N$ cross section a a function of the incident neutrino energy. One clearly observes the enhancement of the $\nu-N$ cross section around $10^6$~GeV if the minimal length is negelcted (indicated as 'without ML')). However, the inclusion of the minimal length effects results in a strong suppression of the cross section enhancement. A very surprising feature is that, despite the high neutrino energies far above the fundamental scale, the charged current cross section remains uninfluenced by the effects of the minimal scale. This fact holds because high momentum transfers are strongly suppressed by the boson propagator in the standard model.

To elaborate these results into realistic predictions for a subsurface neutrino detector such as IceCube, one has to make assumptions about the cosmic neutrino flux $F_\nu(E_\nu)$. For the present study, we have taken the flux from Ref. \cite{Protheroe:1995ft}, which is calculated by considering the propagation of UHE cosmic rays through the extragalactic background radiation field, including interaction-initiated cascades. Further, one has to take into account the geographical situation of a detector, i.e. the screening of the neutrino flux by the surrounding earth. The column density of material between the detector and the upper atmosphere can be approximated by\cite{Morris:1991bb}
$$X(\theta)=\rho\bigg(\sqrt{(R_\oplus-D)^2\cos^2\theta+2DR_\oplus-D^2}-(R_\oplus-D)\cos\theta\bigg)\quad,$$
where $\theta$ is the zenith angle, $\rho$ is the mean earth mass density, $R_\oplus$ is the earth radius and $D$ is the vertical depth of the detector. We have used $D=2{\rm km}$ to obtain realistic predictions for IceCube. With this column density, the number of black hole events per time $t$ and solid angle $\Omega$ with detection threshold energy $E_{th}$ in a subsurface detector with volume $V$ reads
$$\frac{{\rm d^2}N}{{\rm d}t{\rm d}\Omega}=\frac{\rho_{det} V}{M_N}\int_{E_{th}}^\infty {\rm d}E_\nu F_\nu(E_\nu) \sigma_{\nu N\rightarrow BH}(E_\nu)\exp\bigg(-\sigma_{\nu N\rightarrow X} X(\theta)/M_N\bigg)\quad.$$
Here, $\rho_{det}$ is the mass density of the detector material and $\sigma_{\nu N\rightarrow X}$ is the total cross section as a sum of the black hole and the charged current cross section.

The result for the total number of black hole events per year as a function of the fundamental scale $M_f$ and the minimal black hole mass $M_{min}$ is depicted in Fig. \ref{fig3}. As can be seen, the number of black hole events substantially decreases when taking into account effects of a minimal length scale. For a value of $M_f=1{\rm TeV}$ in the range of the electroweak symmetry breaking scale and even with the most optimistic case of $M_{min}=M_f$, there will be basically no black hole events at IceCube. This result is nearly independent of the number of extra dimensions, as can be seen from Fig. \ref{fig4}. A changing number of extra dimensions only slightly affects the number of black holes produced both with and without minimal length scale.

\begin{figure}
\includegraphics[width=14cm]{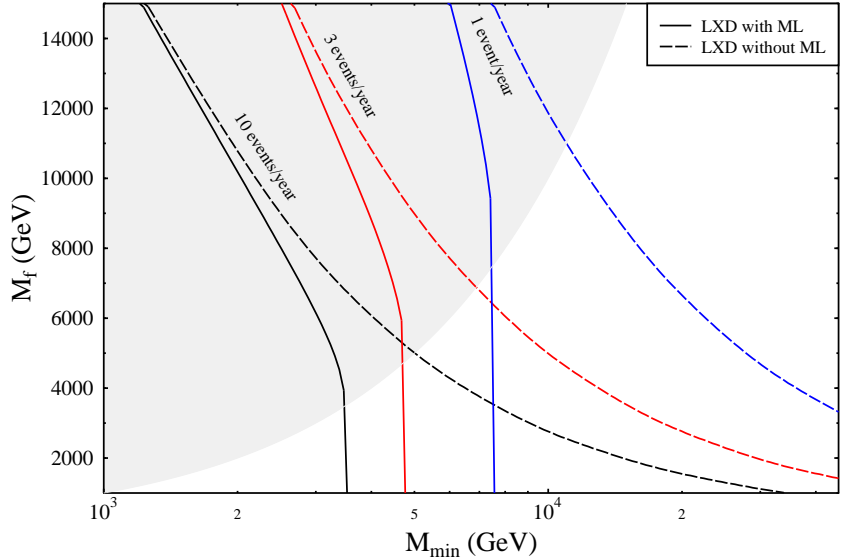}
\caption{Black hole production rates contained in a volume of 1÷km$^3$ at 2÷km below the surface as a function of the fundamental scale and the minimal mass of the black holes. Here, the upper limit to the cosmogenic neutrino flux including hidden sources is assumed \cite{Mannheim:1998wp}. The shaded region is excluded ($M_{min} < M_f$). Again, $d=6$.\label{fig5}}
\end{figure}

Finally, we take a look at the most optimistic parameter set for black hole production, i.e. again a high number of extra dimensions $d=6$, and we assume the maximum possible neutrino flux predicted by hadronic photoproduction models from Ref. \cite{Mannheim:1998wp}. The result is shown in Fig. \ref{fig5}. Due to the substantially larger neutrino flux, the black hole production rates increase, but still the suppression due to the minimal length scale is evident, especially for higher minimal black hole masses $M_{min}$. This is particularly important since $M_{min}$ is supposed to rise in models with a minimal length scale.\cite{Cavaglia:2003qk,Cavaglia:2004jw} Thus, although the here depicted production rates are in the observable region for some section of parameter space, one must keep in mind the very optimistic assumptions that lead to them.

\section{Conclusion}
In summary we have shown, that the inclusion of a minimal length scale motivated by string theory leads to a strong modification of the black hole production cross section in neutrino induced ultra high energy cosmic ray events. It was demonstrated that the strong enhancement of the neutrino-nucleon cross section due to black hole production is severly reduced, in line with the low observation rate of horizontal air showers. In addition, the altered mass distribution of produced black holes results in a strong suppression of the black hole detection rate in neutrino telescopes such as IceCube, questioning previous optimistic estimates that suggest a discovery potential of new physics in such telescopes before the start of LHC.

\section*{Acknowledgments}
U.H. thanks the Frankfurt Institute of Advanced Studies for financial support through a PhD scholarship. This work was supported by GSI and BMBF.


\section*{References}

\begin{thebibliography}{10}

\bibitem{Domokos:1998ry}
  G.~Domokos and S.~Kovesi-Domokos,
  Phys.\ Rev.\ Lett.\  {\bf 82}, 1366 (1999)
  [arXiv:hep-ph/9812260].

\bibitem{Emparan:2001kf}
  R.~Emparan, M.~Masip and R.~Rattazzi,
  Phys.\ Rev.\ D {\bf 65}, 064023 (2002)
  [arXiv:hep-ph/0109287].

\bibitem{Ringwald:2001vk}
  A.~Ringwald and H.~Tu,
  Phys.\ Lett.\ B {\bf 525}, 135 (2002)
  [arXiv:hep-ph/0111042].

\bibitem{Anchordoqui:2001cg}
  L.~A.~Anchordoqui, J.~L.~Feng, H.~Goldberg and A.~D.~Shapere,
  Phys.\ Rev.\ D {\bf 65}, 124027 (2002)
  [arXiv:hep-ph/0112247].

\bibitem{Kazanas:2001ep}
  D.~Kazanas and A.~Nicolaidis,
  Gen.\ Rel.\ Grav.\  {\bf 35}, 1117 (2003)
  [arXiv:hep-ph/0109247].

\bibitem{Kowalski:2002gb}
  M.~Kowalski, A.~Ringwald and H.~Tu,
  Phys.\ Lett.\ B {\bf 529}, 1 (2002)
  [arXiv:hep-ph/0201139].

\bibitem{Alvarez-Muniz:2002ga}
  J.~Alvarez-Muniz, J.~L.~Feng, F.~Halzen, T.~Han and D.~Hooper,
  Phys.\ Rev.\ D {\bf 65}, 124015 (2002)
  [arXiv:hep-ph/0202081].

\bibitem{Sigl:2002bb}
  G.~Sigl,
  arXiv:hep-ph/0207254.

\bibitem{Anchordoqui:2003jr}
  L.~A.~Anchordoqui, J.~L.~Feng, H.~Goldberg and A.~D.~Shapere,
  Phys.\ Rev.\ D {\bf 68}, 104025 (2003)
  [arXiv:hep-ph/0307228].

\bibitem{Anchordoqui:2004ma}
  L.~A.~Anchordoqui, Z.~Fodor, S.~D.~Katz, A.~Ringwald and H.~Tu,
  JCAP {\bf 0506}, 013 (2005)
  [arXiv:hep-ph/0410136].

\bibitem{Anchordoqui:2005gj}
  L.~A.~Anchordoqui, H.~Goldberg, M.~C.~Gonzalez-Garcia, F.~Halzen, D.~Hooper, S.~Sarkar and T.~J.~Weiler,
  Phys.\ Rev.\ D {\bf 72}, 065019 (2005)
  [arXiv:hep-ph/0506168].

\bibitem{Anchordoqui:2005is}
  L.~Anchordoqui and F.~Halzen,
  arXiv:hep-ph/0510389.

\bibitem{Arkani-Hamed:1998rs}
  N.~Arkani-Hamed, S.~Dimopoulos and G.~R.~Dvali,
  Phys.\ Lett.\ B {\bf 429}, 263 (1998)
  [arXiv:hep-ph/9803315].

\bibitem{Antoniadis:1998ig}
  I.~Antoniadis, N.~Arkani-Hamed, S.~Dimopoulos and G.~R.~Dvali,
  Phys.\ Lett.\ B {\bf 436}, 257 (1998)
  [arXiv:hep-ph/9804398].

\bibitem{Feng:2001ib}
  J.~L.~Feng and A.~D.~Shapere,
  Phys.\ Rev.\ Lett.\  {\bf 88}, 021303 (2002)
  [arXiv:hep-ph/0109106].

\bibitem{Jain:2000pu}
  P.~Jain, D.~W.~McKay, S.~Panda and J.~P.~Ralston,
  Phys.\ Lett.\ B {\bf 484}, 267 (2000)
  [arXiv:hep-ph/0001031].

\bibitem{Anchordoqui:2001ei}
  L.~Anchordoqui and H.~Goldberg,
  Phys.\ Rev.\ D {\bf 65}, 047502 (2002)
  [arXiv:hep-ph/0109242].

\bibitem{Ahn:2003qn}
  E.~J.~Ahn, M.~Ave, M.~Cavaglia and A.~V.~Olinto,
  Phys.\ Rev.\ D {\bf 68}, 043004 (2003)
  [arXiv:hep-ph/0306008].

\bibitem{Ave:2003ew}
  M.~Ave, E.~J.~Ahn, M.~Cavaglia and A.~V.~Olinto,
  arXiv:astro-ph/0306344.

\bibitem{Tyler:2000gt}
  C.~Tyler, A.~V.~Olinto and G.~Sigl,
  Phys.\ Rev.\ D {\bf 63}, 055001 (2001)
  [arXiv:hep-ph/0002257].

\bibitem{Anchordoqui:2005ey}
  L.~Anchordoqui, T.~Han, D.~Hooper and S.~Sarkar,
  arXiv:hep-ph/0508312.

\bibitem{Uehara:2001yk}
  Y.~Uehara,
  Prog.\ Theor.\ Phys.\  {\bf 107}, 621 (2002)
  [arXiv:hep-ph/0110382].

\bibitem{Ahn:2003cz}
  E.J. Ahn, M. Cavaglia and A.V. Olinto,
  Astropart.\ Phys.\ {\bf 22}, 377 (2005) 
  [arXiv:hep-ph/0312249].

\bibitem{Cavaglia:2003qk}
  M.~Cavaglia, S.~Das and R.~Maartens,
  Class.\ Quant.\ Grav.\  {\bf 20}, L205 (2003)
  [arXiv:hep-ph/0305223].

\bibitem{Cavaglia:2004jw}
  M.~Cavaglia and S.~Das,
  Class.\ Quant.\ Grav.\  {\bf 21}, 4511 (2004)
  [arXiv:hep-th/0404050].

\bibitem{Hossenfelder:2004ze}
  S.~Hossenfelder,
  Phys.\ Lett.\ B {\bf 598}, 92 (2004)
  [arXiv:hep-th/0404232].

\bibitem{Gandhi:1995tf}
  R.~Gandhi, C.~Quigg, M.~H.~Reno and I.~Sarcevic,
  Astropart.\ Phys.\  {\bf 5}, 81 (1996)
  [arXiv:hep-ph/9512364].

\bibitem{Argyres:1998qn}
  P.~C.~Argyres, S.~Dimopoulos and J.~March-Russell,
  Phys.\ Lett.\ B {\bf 441}, 96 (1998)
  [arXiv:hep-th/9808138].

\bibitem{Myers:1986un}
  R.~C.~Myers and M.~J.~Perry,
  Annals Phys.\  {\bf 172}, 304 (1986).

\bibitem{Gross:1988ar}
  D.J. Gross and P.F. Mende,
  Nucl.\ Phys.\ B {\bf 303}, 407 (1988).

\bibitem{Amati:1988tn}
  D.~Amati, M.~Ciafaloni and G.~Veneziano,
  Phys.\ Lett.\ B {\bf 216}, 41 (1989).

\bibitem{Witten:1997fz}
  E.~Witten,
  Phys.\ Today {\bf 50N5}, 28 (1997).

\bibitem{Hossenfelder:2003jz}
  S.~Hossenfelder, M.~Bleicher, S.~Hofmann, J.~Ruppert, S.~Scherer and H.~Stoecker,
  Phys.\ Lett.\ B {\bf 575}, 85 (2003)
  [arXiv:hep-th/0305262].

\bibitem{Hossenfelder:2004up}
  S.~Hossenfelder,
  Phys.\ Rev.\ D {\bf 70}, 105003 (2004)
  [arXiv:hep-ph/0405127].

\bibitem{Ahluwalia:2000iw}
  D.~V.~Ahluwalia,
  Phys.\ Lett.\ A {\bf 275}, 31 (2000)
  [arXiv:gr-qc/0002005].

\bibitem{Kempf:1995su}
  A. Kempf, G. Mangano and R.B. Mann,
  Phys.\ Rev.\ D {\bf 52}, 1108 (1995) 1108 
  [arXiv:hep-th/9412167].

\bibitem{Protheroe:1995ft}
  R.~J.~Protheroe and P.~A.~Johnson,
  Astropart.\ Phys.\  {\bf 4}, 253 (1996)
  [arXiv:astro-ph/9506119].

\bibitem{Morris:1991bb}
  D.~A.~Morris and R.~Rosenfeld,
  Phys.\ Rev.\ D {\bf 44}, 3530 (1991).

\bibitem{Mannheim:1998wp}
  K.~Mannheim, R.~J.~Protheroe and J.~P.~Rachen,
  Phys.\ Rev.\ D {\bf 63}, 023003 (2001)
  [arXiv:astro-ph/9812398].


\end{thebibliography}
\end{document}